\documentclass[12pt]{article}

\usepackage{amssymb}
\usepackage{amsmath}
\usepackage{amsfonts}

\newcommand{\be}{\begin{equation}}
\newcommand{\ee}{\end{equation}}
\newcommand{\bea}{\begin{eqnarray}}
\newcommand{\eea}{\end{eqnarray}}

\newcommand{\kt}{\rangle}
\newcommand{\br}{\langle}

\newcommand{\ed}{\end{document}}

\begin{document}

\title{Comment on Complex Extension of Quantum Mechanics}
\author{Ali Mostafazadeh\thanks{E-mail address:
amostafazadeh@ku.edu.tr}\\ {\small Department of Mathematics,
Ko\c{c} University, 34450 Sariyer, Istanbul, Turkey}}
\date{ }
\maketitle

\begin{abstract}
In their Erratum [Phys.\ Rev.\ Lett.\ {\bf 92}, 119902 (2004),
quant-ph/0208076], written in reaction to [quant-ph/0310164],
Bender, Brody and Jones propose a revised definition for a
physical observable in ${\cal PT}$-symmetric quantum mechanics. We
show that although this definition avoids the dynamical
inconsistency revealed in quant-ph/0310164, it is still not a
physically viable definition. In particular, we point out that a
general proof that this definition is consistent with the
requirements of the quantum measurement theory is lacking, give
such a proof for a class of ${\cal PT}$-symmetric systems by
establishing the fact that this definition implies that the
observables are pseudo-Hermitian operators, and show that for all
the cases that this definition is consistent with the requirements
of measurement theory it reduces to a special case of a more
general definition given in [quant-ph/0310164]. The latter is the
unique physically viable definition of observables in ${\cal
PT}$-symmetric quantum mechanics.
\end{abstract}

Bender, Brody and Jones [Phys.\ Rev.\ Lett.\ {\bf 92}, 119902
(2004), quant-ph/0208076] have recently proposed the following
definition of a physical observable in ${\cal PT}$-symmetric
quantum mechanics. {\bf Def.~1:} {\em A linear operator $A$ is
called an observable if it satisfies} \cite{f}
    \be
    A^T={\cal CPT} A\: {\cal CPT}.
    \label{d3}
    \ee
This definition avoids the incompatibility of their initial
definition \cite{bbj} with the dynamical aspects of the theory
\cite{c1}. The purpose of this comment letter is to use the
requirements of the quantum measurement theory to provide a
critical assessment of the viability of Def.~1. In particular, we
point out that (a) a general proof that Def.~1 is consistent with
these requirements is lacking, (b) give such a proof for a class
of ${\cal PT}$-symmetric systems by establishing the fact that
(\ref{d3}) implies that $A$ is a pseudo-Hermitian operator
\cite{p1}, and (c) show that for all the cases that Def.~1 is
consistent with these requirements it reduces to a more general
definition \cite{p58}, namely {\bf Def.~2:} {\em A linear operator
$A$ is called an observable if it is Hermitian with respect to the
${\cal CPT}$-inner product $\br\cdot|\cdot\kt$, i.e.,
$\br\cdot|A\cdot\kt=\br A\cdot|\cdot\kt$}.

Standard quantum measurement theory imposes the following
conditions on any linear operator $A$ that is to be identified
with a physical observable. {\bf (i)} the eigenvalues of $A$ must
be real; {\bf (ii)} $A$ has a complete set of eigenvectors that
are mutually orthogonal with respect to the defining inner product
$\br\cdot|\cdot\kt$ of the Hilbert space ${\cal H}$.

It is a well-known result of linear algebra that (i) and (ii) are
the necessary and sufficient conditions for the Hermiticity of an
operator $A$, i.e., $\br\cdot|A\cdot\kt=\br A\cdot|\cdot\kt$. In
${\cal PT}$-symmetric QM, $\br\cdot|\cdot\kt$ is the ${\cal
CPT}$-inner product \cite{bbj}. This shows that the most general
definition that is compatible with (i) and (ii) is Def.~2. As a
result, Def.~1 would be a physically viable definition, only if it
turns out to be a special case of Def.~2. It is not equivalent to
Def.~2, for it puts the additional restriction that the
Hamiltonian $H$ be not only ${\cal PT}$-symmetric but also
symmetric ($H^T=H$); it cannot for example be used to determine
the observables for the ${\cal PT}$-symmetric system defined by
the Hamiltonian $H=(p+ix)^2+x^2$, \cite{jpa-2003}.

Next, we note that one can use the identities $[{\cal P},{\cal
T}]=[{\cal C},{\cal PT}]=0$ and ${\cal C}^2={\cal P}^2=1$ to show
that Eq.~(\ref{d3}) implies
    \be
    A^\dagger=\eta_+^{-1} A\,\eta_+,
    \label{ph}
    \ee
where $A^\dagger={\cal T}A^T{\cal T}$ is the usual adjoint of $A$
and $\eta_+:={\cal PC}$. Eq.~(\ref{ph}) is the defining relation
for a pseudo-Hermitian operator \cite{p1}. It is equivalent to the
condition that $A$ be Hermitian with respect to the inner product
$\br\cdot,\cdot\kt_{\eta_+}:=(\cdot,\eta_+\cdot)$ where
$(\cdot,\cdot)$ is the ordinary $L^2$-inner product. Therefore,
Def.~1 implies that the observables $A$ are Hermitian operators
with respect to $\br\cdot,\cdot\kt_{\eta_+}$, i.e.,
$\br\cdot,A\cdot\kt_{\eta_+}=\br A\cdot,\cdot\kt_{\eta_+}$. For
${\cal PT}$-symmetric theories defined on the real line, one can
show by a direct computation \cite{jmp-2003} that
$\br\cdot,\cdot\kt_{\eta_+}$ coincides with the ${\cal CPT}$-inner
product. This proves that for these theories Def.~1 does indeed
adhere to the requirements (i) and (ii) above. For ${\cal
PT}$-symmetric theories defined using a complex contour, such a
proof is lacking.

This is a serious shortcoming. In effect it means that in order to
employ Def.~1 one must not only establish the reality of the
eigenvalues of an observable $A$ but also prove that (\ref{d3})
implies the completeness of the eigenvectors of $A$ and their
orthogonality. Moreover, Def.~1 does not provide any practical
means to construct the observables of the theories to which it
applies. As argued in \cite{p58} the situation is different if one
adopts Def.~2. One then would just compute the matrix elements
$A_{mn}=\br\phi_m|A\phi_n\kt$ in the energy eigenbasis
$\{\phi_n\}$ and check whether $A_{mn}^*=A_{nm}$.

In conclusion, there is no logical reason why one should adopt
Def.~1 while there is already an alternative, namely Def.~2, that
avoids all the above-mentioned problems. A conceptual consequence
of adapting Def.~2 is that the only structural difference between
conventional QM and ${\cal PT}$-symmetric QM is that in the latter
one defines the Hilbert space using the eigenvalue problem of a
differential operator. As explained in \cite{p58} the fact that
there is (up to unitary equivalence) a single separable Hilbert
space shows that this difference does not have any fundamental
ramifications. This in turn suggests that the ${\cal
PT}$-symmetric QM should be viewed as a framework for dealing with
phenomenological and effective theories.
$$
***************************
$$
This work has been supported by the Turkish Academy of Sciences in
the framework of the Young Researcher Award Program
(EA-T$\ddot{\rm U}$BA-GEB$\dot{\rm I}$P/2001-1-1).

{\small

\ed
\begin{thebibliography}{9}
\bibitem{f} According to \cite{bbj}, given a linear operator $A$
represented by the infinite matrix $A(x,x')$ in the standard
position representation, the traspose $A^T$ of $A$ is the operator
that is represented by $A^T(x,x'):=A(x',x)$. $A$ is called
symmetric if $A^T=A$.
\bibitem{c1} This incompatibility was initially revealed in an
attempt to construct the observables for a finite-dimensional toy
model. It was then generally established in \cite{p58}.
\end{thebibliography}
